\documentclass[twocolumn,showpacs,preprintnumbers,amsmath,amssymb]{revtex4-1}

\usepackage{epsfig}
\usepackage{graphicx}
\usepackage{dcolumn}
\usepackage{bm} 
\usepackage{subfigure}
\usepackage{appendix}

\newcommand{\beq}{\begin{equation}}
\newcommand{\eeq}{\end{equation}}
\newcommand{\beqa}{\begin{eqnarray}}
\newcommand{\eeqa}{\end{eqnarray}}

\newcommand{\la}{\langle}
\newcommand{\ra}{\rangle}
\newcommand{\ket}[1]{|#1\rangle}                
\newcommand{\bra}[1]{\langle#1|}                
\newcommand{\scpr}[2]{\langle#1|#2\rangle}      
\newcommand{\matel}[3]{\langle#1|#2|#3\rangle}  

\def\oc#1{{ Opt.\ Commun.} {\bf#1}}

\def\jpa#1{{ J.\ Phys.\ A} {\bf#1}}

\def\pra#1{{ Phys.\ Rev. A\/} {\bf#1}}
\def\prb#1{{ Phys.\ Rev. B\/} {\bf#1}}

\def\prl#1{{ Phys.\ Rev.\ Lett.} {\bf#1}}

\begin{document}

\title{Two Qubits Tavis-Cummings Model Beyond the Rotating Wave Approximation: Degenerate Regime}

\author{S. Agarwal, S.M. Hashemi Rafsanjani and J.H. Eberly}

\affiliation{ Rochester Theory Center and the Department of Physics
\& Astronomy\\
University of Rochester, Rochester, New York 14627}

\email{shantanu@pas.rochester.edu}


\date{\today}

\begin{abstract}
We study the dynamics of two qubits interacting with a single mode of a harmonic oscillator beyond the rotating wave approximation in the ideally degenerate regime. Exact analytic expressions are obtained for state properties of interest, including qubit entanglement for a certain class of initial states of the oscillator and the qubits. Qualitative differences and similarities in the evolution of the qubits in the degenerate regime when the oscillator is treated quantum mechanically and classically are discussed.
\end{abstract}

\pacs{..........}


\maketitle




\section{Introduction}
Two level systems that interact with a harmonic oscillator can model many physical phenomena, such as atoms interacting 
with an electromagnetic field \cite{JC, Allen-Eberly}, electrons coupled to a phonon mode of a crystal lattice \cite{Holstein}, or super-conducting qubits interacting with a nano-mechanical resonator \cite{Irish-Schwab, Schwab-Roukes}, a transmission line resonator \cite{Blais-etal, Wallraff-etal}, or an LC circuit \cite{Chiorescu-etal, Johansson-etal}. In the area of quantum optics, the model describing the interaction of a single qubit with a harmonic oscillator, called the Jaynes-Cummings model (JCM) \cite{JC} is extensively studied in the literature \cite{Shore-Knight}. A generalization of the JCM to include many qubits not interacting with each other, but interacting with a single mode of a quantized harmonic oscillator, called the 
Tavis-Cummings model (TCM) \cite{Tavis-Cummings-1, Tavis-Cummings-2}, is also of great interest as it provides a framework to study collective properties of the qubits. 

In physical situations where the qubits are nearly resonant with the oscillator and the coupling between the qubits and the oscillator is weak, it is a good approximation to drop certain terms, called the counter-rotating terms, from the Hamiltonian describing the evolution of the qubits and the oscillator. Under this approximation, called the rotating wave approximation (RWA), the dynamics of the system can be solved analytically \cite{JC}. 

In optical setups, the near resonance and the weak coupling conditions are usually satisfied under appropriate conditions, justifying the use of the RWA to describe the dynamics of the system \cite{Allen-Eberly}. With the recent advancements in the area of circuit QED, it is now possible to engineer systems for which the near resonance and/or the weak coupling conditions are not satisfied \cite{Niemczyk-etal, Forn-etal, Fedorov-etal}. To understand the dynamics under such conditions, it is necessary to study the evolution of the system beyond the RWA. Analytical approximations and numerical methods have been used to understand both qualitative and quantitative features of the dynamics  beyond the RWA for the Jaynes-Cummings \cite{Swain, Zaheer-Zubairy, Phoenix, Shore-book, Crisp-92, Finney-Banacloche, Tur-2000, Tur-2001, Irish-05, Irish-07, Larson, Hausinger-08, Hausinger-10, Casanova-etal} and the Tavis-Cummings \cite{Seke-1, Seke-2, Klimov-Chumakov, Zueco-etal} model. 

In this report, we study quantum collective dynamics, in particular the entanglement properties of the two qubits TCM, beyond the RWA, in the regime when both the qubits are far from resonance with the oscillator. We extend recent treatments \cite{Jing-Ficek, Ficek_etal, Chen_etal, Rafsanjani_etal} of the role of the counter-rotating terms in the dynamics of entanglement between the qubits.

In particular, we examine the idealized case of zero-gap qubits, focusing on the entanglement dynamics of the four Bell states with the oscillator initially in either a thermal state, a coherent state or a number state. A study of the dynamics of a degenerate qubit interacting with a classical field, and a discussion of a physical system which can be treated as a degenerate qubit, has been given by Shakov and McGuire \cite{Shakov-McGuire}.

We find the eigenvalues and eigenfunctions of the Hamiltonian of a two-qubit TCM in the degenerate regime in section \ref{s.model}. In section \ref{s.dynamics}, using the Glauber-Sudarshan \textit{P} representation \cite{Glauber, Sudarshan} to describe the initial state of the oscillator, we find the evolution of any initial product state of the qubits and the oscillator. The entanglement properties between the qubits for a certain class of initial states are studied in section \ref{s.dynamics_ent}. In section 
\ref{s.comparison}, it is shown that the dynamics of the qubits is qualitatively different for the quantum mechanical and the classical descriptions of the oscillator and we conclude in section \ref{s.conclusion}.




\section{The Model}\label{s.model}

The Hamiltonian describing the dynamics of two qubits 
interacting with a single harmonic oscillator is given by \cite{Tavis-Cummings-1}:
\beq\label{Hamiltonian}
\hat{H}=\hbar\frac{\omega_{0}}{2}(\hat{\sigma}_{z}^{(1)}+\hat{\sigma}_{z}^{(2)}) + \hbar\omega \hat{a}^{\dagger}\hat{a}
+ \hbar\lambda(\hat{a}+\hat{a}^{\dagger})(\hat{\sigma}_{x}^{(1)}+\hat{\sigma}_{x}^{(2)}).
\eeq
The two qubits are assumed to be similar but fundamentally distinguishable.
The transition frequency of each qubit (assumed to be the same for simplicity)
is $\omega_{0}$ and the oscillator frequency is $\omega$. The coupling constant between oscillator and both the qubits (again assumed to be the same for mathematical convenience) is $\lambda$. The raising and lowering operators for the harmonic oscillator are conventional. The qubit operators $\hat{\sigma}_{z}^{(i)}$ and $\hat{\sigma}_{x}^{(i)}$ are the usual Pauli matrices in the Hilbert space of the $i^{th}$ qubit. The interaction part of the Hamiltonian that couples the qubits to the oscillator is $\hat{H}_{I}=\hbar\lambda(\hat{a}+\hat{a}^{\dagger})(\hat{\sigma}_{x}^{(1)}+\hat{\sigma}_{x}^{(2)})$.

\begin{figure}
\includegraphics[width= 6cm]{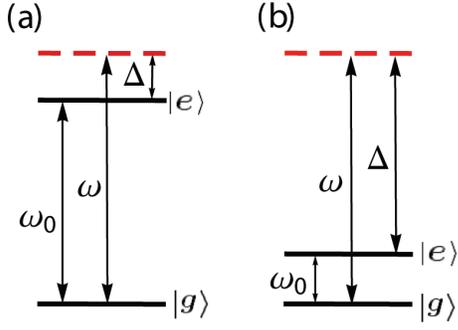}
\caption{Energy level diagrams for: (a) $\omega_0\approx\omega$ and (b) $\omega_0\ll\omega$.} \label{f.model}
\end{figure}

While discussing the RWA, it is suggestive to write the interaction Hamiltonian as:
\beq\label{H_I}
\hat{H}_{I}=\hbar\lambda(\hat{a}+\hat{a}^{\dagger})(\hat{\sigma}_{+}^{(1)}+\hat{\sigma}_{-}^{(1)})+
\hbar\lambda(\hat{a}+\hat{a}^{\dagger})(\hat{\sigma}_{+}^{(2)}+\hat{\sigma}_{-}^{(2)}).
\eeq
If
$\lambda\ll\omega,\omega_{0}$ and $\omega_{0}\approx\omega$, one can neglect the
counter rotating terms: $\hat{a}^{\dagger}\hat{\sigma}_{+}^{(i)}$ and $\hat{a}\hat{\sigma}_{-}^{(i)}$. 
This is called the rotating wave approximation \cite{JC,Allen-Eberly}. The condition $\omega_{0}\approx\omega$, when the RWA is valid is shown in Fig. \ref{f.model}(a). The zero-gap degenerate regime that we are interested in takes $\omega_{0}=0$, an idealization of the case when $\omega_{0}\ll\omega$ and $\omega_{0}\ll\lambda$. The condition $\omega_{0}\ll\omega$ is shown in Fig. \ref{f.model}(b).

When $\omega_0=0$, the non-RWA Hamiltonian takes the form 
\beq\label{e.H_o}
\hat{H}_{0}=\hbar\omega \hat{a}^{\dagger}\hat{a}
+ \hbar\lambda(\hat{a}+\hat{a}^{\dagger})(\hat{\sigma}_{x}^{(1)}+\hat{\sigma}_{x}^{(2)}).
\eeq
In order to study the dynamics of the system in the degenerate regime, 
we first find the eigenstates, $\ket{\Phi}$, and eigenvalues, $E$, of $\hat{H}_0$:
\beq\label{e.Degen_Eigen1}
\left[\hbar\omega \hat{a}^{\dagger}\hat{a}
+ \hbar\lambda(\hat{a}+\hat{a}^{\dagger})(\hat{\sigma}_{x}^{(1)}+\hat{\sigma}_{x}^{(2)})\right]\ket{\Phi}=E\ket{\Phi}.
\eeq
Eigenstates $\ket{\Phi}$ will be of the form $\ket{j,m}\ket{\phi_{m}}$ where $\ket{j,m}$ are the eigenstates of 
$(\hat{\sigma}_{x}^{(1)}+\hat{\sigma}_{x}^{(2)})$ and $\ket{\phi_{m}}$ are the oscillator eigenstates found from 
$\hat{H}_0$ by replacing $(\hat{\sigma}_{x}^{(1)}+\hat{\sigma}_{x}^{(2)})$ by the eigenvalue corresponding to $\ket{j,m}$ \cite{Irish-05}. 

The four eigenstates of $(\hat{\sigma}_{x}^{(1)}+\hat{\sigma}_{x}^{(2)})$ are:
\beq
\ket{j,m}=\ket{1,\pm1}\mbox{, }\ket{1,0}\mbox{ and }\ket{0,0},
\eeq
with eigenvalues $2m$.
In terms of the simultaneous eigenstates of $\hat{\sigma}_x^{(1)}$ and $\hat{\sigma}_x^{(2)}$, 
$\hat{\sigma}_x^{(i)}\ket{\pm}=\pm\ket{\pm}$, the states $\ket{j,m}$ can be written as:
\begin{equation}\label{e.unit_trans}
 \begin{pmatrix} \ket{1,1} \\ \ket{1,0} \\ \ket{0,0} \\ \ket{1,-1}\end{pmatrix}=
 \begin{pmatrix} 1 & 0 & 0 & 0 \\ 
                 0 & 1/\sqrt{2} & 1/\sqrt{2} & 0 \\
                 0 & 1/\sqrt{2} & -1/\sqrt{2} & 0 \\
                 0 & 0 &  0 & 1 \\
 \end{pmatrix}
 \begin{pmatrix} \ket{+,+} \\ \ket{+,-} \\ \ket{-,+} \\ \ket{-,-}\end{pmatrix}.
\end{equation}

Having found $\ket{j,m}$, let us now find $\ket{\phi_{m}}$ that satisfy the eigenvalue equation: 
\beq\label{e.Degen_Eigen2}
\left[\hbar\omega \hat{a}^{\dagger}\hat{a}
+ \hbar2m\lambda(\hat{a}+\hat{a}^{\dagger})\right]\ket{\phi_{m}}=E\ket{\phi_{m}}.
\eeq  
We denote $\lambda/\omega$ by $\beta$ and $m\lambda/\omega$ by $\beta_m$. Taking $\beta_m$ to be real, and completing the square in equation 
(\ref{e.Degen_Eigen2}), we get:
\beq\label{e.Degen_Eigen3}
(\hat{a}^{\dagger}+2\beta_m)(\hat{a}+2\beta_m)\ket{\phi_m}
=\left(\frac{E}{\hbar\omega}+4\beta^2_m\right)\ket{\phi_m}.
\eeq  
Using the displacement operator, $\hat{D}(x)=\exp[x(\hat{a}^{\dagger}-\hat{a})]$ (for real $x$), we can write the expression on the left hand of (\ref{e.Degen_Eigen3}) as:
\beq\label{e.Degen_Eigen4}
(\hat{a}^{\dagger}+2\beta_m)(\hat{a}+2\beta_m)\ket{\phi_m}
=\hat{D}(-2\beta_m)\hat{a}^{\dagger}\hat{a}\hat{D}(2\beta_m)\ket{\phi_m}.
\eeq
As the operator $\hat{D}(-2\beta_m)\hat{a}^{\dagger}\hat{a}\hat{D}(2\beta_m)$ is the number operator for a displaced harmonic oscillator with eigenstates $\hat{D}(-2\beta_m)\ket{N}$, we get from equations (\ref{e.Degen_Eigen3}) and (\ref{e.Degen_Eigen4}) that
\beq
\ket{\phi_m}=\hat{D}(-2\beta_m)\ket{N}\equiv\ket{N_m}.
\eeq
Here $\ket{N}$ is the number state of the undisplaced oscillator, i.e. $\hat{a}^{\dagger}\hat{a}\ket{N}=N\ket{N}$.

Corresponding to the state $\ket{N_m}$, the energy, $E$, in equation (\ref{e.Degen_Eigen2}) takes value:
\beq
E_{N,m}=\hbar\omega(N-4\beta^2_m).
\eeq
In the discussion of two level systems interacting with a harmonic oscillator beyond the RWA, the use of a displaced harmonic oscillator basis was first used by Crisp \cite{Crisp-92}. The displaced oscillator states have the properties:
\beqa
\scpr{N_{m}}{N'_{m}}&=&\delta_{N,N'},\nonumber\\
\scpr{N_{m}}{M_{m'}}&\neq&0.
\eeqa
The non-orthogonality condition, $\scpr{N_{m}}{M_{m'}}\neq0$, plays an important role when one discusses the spectrum of the system for $\omega_0\neq0$.   




\section{Evolution of any initial product state of oscillator and qubits}\label{s.dynamics}

Let the initial state of the system be given by the density matrix $\hat{\rho}(0)=\hat{Q}\otimes\hat{F}$, where $\hat{Q}$ is the joint density matrix of both the qubits and $\hat{F}$ is the density matrix of the oscillator. Writing $\hat{Q}$ in the basis $\ket{j,m}$, and the oscillator density matrix in the diagonal coherent state basis (Glauber-Sudarshan $\it{P}$ representation), $\hat{\rho}(0)$ takes the form:
\beq\label{e.initialstate}
\hat{\rho}(0)=\sum Q_{jm;j'm'}\ket{j,m}\bra{j',m'}\otimes
\int \mathrm{d^2}\alpha P(\alpha)\ket{\alpha}\bra{\alpha},
\eeq
where $Q_{jm;j'm'}=\matel{j,m}{\hat{Q}}{j',m'}$. Evolving $\hat{\rho}(0)$ in time, we get:
\beqa\label{e.rhotformal}
\hat{\rho}(t)&=&e^{-i\hat{H}_0 t/\hbar}\hat{\rho}(0)e^{i\hat{H}_0 t/\hbar},\nonumber\\
&=&\sum \int \mathrm{d^2}\alpha P(\alpha) Q_{jm;j'm'}\nonumber\\
&&\times e^{-i\hat{H}_0 t/\hbar}\ket{j,m}\ket{\alpha}\bra{\alpha}\bra{j',m'}e^{i\hat{H}_0 t/\hbar}.
\eeqa 
From Appendix A, we get
\beqa\label{e.termevol}
e^{-i\hat{H}_0 t/\hbar}\ket{j,m}\ket{\alpha}&=&\ket{j,m}\ket{(\alpha+2\beta_m)e^{-i\omega t}-2\beta_m}\nonumber\\
&\times&\exp{\left(-2i\beta_m f(\omega t,\alpha)\right)}\nonumber\\
&\times&\exp{\left(4i\beta_m^2(\omega t-\sin{\omega t})\right)},
\eeqa
where
\beq
f(\omega t,\alpha)=\sin{\frac{\omega t}{2}}(\alpha^*e^{i\omega t/2}+\alpha e^{-i\omega t/2}).
\eeq
Using equation (\ref{e.termevol}), $\hat{\rho}(t)$ takes the form:
\beqa\label{e.rhot}
\hat{\rho}(t)&=&\sum \int \mathrm{d^2}\alpha P(\alpha) Q_{jm;j'm'}\nonumber\\
&\times&\ket{j,m}\bra{j',m'}\nonumber\\
&\otimes&\ket{(\alpha+2\beta_m)e^{-i\omega t}-2\beta_m}
\bra{(\alpha+2\beta_{m'})e^{-i\omega t}-2\beta_{m'}}\nonumber\\
&\times&\exp{\left(-2i(\beta_m-\beta_{m'})f(\omega t,\alpha)\right)}\nonumber\\
&\times&\exp{\left(4i(\beta_m^2-\beta_{m'}^2)(\omega t-\sin{\omega t})\right)}.
\eeqa 
The dynamics of the qubits can be followed by tracing the oscillator degrees of freedom from equation (\ref{e.rhot}) to get the reduced density matrix for the qubits:
\beqa
\hat{\rho}_q(t)&=&\mbox{Tr}_o\{\hat{\rho}(t)\},\nonumber\\ 
&=&\sum \int \mathrm{d^2}\alpha P(\alpha) Q_{jm;j'm'}\nonumber\\
&\times&\ket{j,m}\bra{j',m'}\nonumber\\
&\times&\scpr{(\alpha+2\beta_{m'})e^{-i\omega t}-2\beta_{m'}}{(\alpha+2\beta_m)e^{-i\omega t}-2\beta_m}\nonumber\\
&\times&\exp{\left(-2i(\beta_m-\beta_{m'})f(\omega t,\alpha)\right)}\nonumber\\
&\times&\exp{\left(4i(\beta_m^2-\beta_{m'}^2)(\omega t-\sin{\omega t})\right)}.
\eeqa
Using the relation of scalar product between two coherent states, $\scpr{\nu_2}{\nu_1}=e^{-|\nu_1-\nu_2|^2/2}e^{(\nu_2^*\nu_1-\nu_2\nu_1^*)/2}$, we get:
\beqa\label{e.qubit_dens_mat}
\hat{\rho}_q(t)&=&\sum \int \mathrm{d^2}\alpha P(\alpha) Q_{jm;j'm'}\nonumber\\
&\times&\ket{j,m}\bra{j',m'}\nonumber\\
&\times&\exp{\left(-8(\beta_m-\beta_{m'})^{2}\sin^{2}{\omega t/2}\right)}\nonumber\\
&\times&\exp{\left(-4i(\beta_m-\beta_{m'})f(\omega t,\alpha)\right)}\nonumber\\
&\times&\exp{\left(4i(\beta_m^2-\beta_{m'}^2)(\omega t-\sin{\omega t})\right)}.
\eeqa
The reduced density matrix for the qubits, equation (\ref{e.qubit_dens_mat}), will be used in the next section for studying the dynamics of entanglement between the two non-interacting qubits.

It must be noted that for some oscillator states, the function $P(\alpha)$ diverges, e.g. squeezed state. In such cases, the integral in equation (\ref{e.qubit_dens_mat}) can be ill-defined. To deal with such states, instead of using the 
$P(\alpha)$ representation, one might find it convenient to write $\hat{F}$ as:
\beq
\hat{F}=\frac{1}{\pi^2}\int \mathrm{d^2}\alpha\mathrm{d^2}\alpha' \ket{\alpha}\matel{\alpha}{\hat{F}}{\alpha'}\bra{\alpha'}
\eeq
and then follow the evolution of the composite system.  



\section{Dynamics of entanglement between the qubits}\label{s.dynamics_ent}
Using a Hamiltonian similar to (\ref{e.H_o}), it was shown in \cite{Oh-Kim} that two non-interacting, initially unentangled qubits can get entangled due to their interaction with a common bath (harmonic oscillator) initially in the ground state. 
In this section, we study the dynamics of entanglement between the two qubits for various different initial states of the oscillator and the qubits. For all the initial states that we study, the qubits and the oscillator are chosen to be initially separable, and are taken to be of the form of equation (\ref{e.initialstate}). Of all the possible initial states of the form (\ref{e.initialstate}), we choose to analyze the states for which the qubits are initially maximally entangled and are in one of the four possible Bell states:
\beqa
\ket{\Psi_{\pm}}&=&\frac{1}{\sqrt{2}}(\ket{e,g}\pm\ket{g,e}),\nonumber\\
\ket{\Phi_{\pm}}&=&\frac{1}{\sqrt{2}}(\ket{e,e}\pm\ket{g,g}),
\eeqa
where $\hat{\sigma}_z^{(i)}\ket{e}=\ket{e}$ and $\hat{\sigma}_z^{(i)}\ket{g}=-\ket{g}$.

From equation (\ref{e.unit_trans}), we know that for any initial state of the oscillator, the states $\ket{\Psi_{-}}$ and $\ket{\Phi_{-}}$ are eigenstates of $\hat{H}_0$. So, the reduced density matrices for the qubits will not change if they start in the state $\ket{\Psi_{-}}$ or $\ket{\Phi_{-}}$ and the qubits will remain maximally entangled no matter what the initial state of the oscillator is.  

The local unitary operator, $\left(\hat{\sigma}_{x}^{(1)}\otimes\hat{1}^{(2)}\right)$, commutes with the Hamiltonian and connects the states $\ket{\Psi_{+}}$ and $\ket{\Phi_{+}}$:
\beqa
&&\ket{\Psi_{+}}=\left(\hat{\sigma}_{x}^{(1)}\otimes\hat{1}^{(2)}\right)\ket{\Phi_{+}},\nonumber\\
&&\left[\hat{\sigma}_{x}^{(1)}\otimes\hat{1}^{(2)},\hat{H}_{0}\right]=0. 
\eeqa
Since any measure of entanglement calculated for two different states that are connected by a local unitary transformation has the same value \cite{Horedecki_clan}, the entanglement dynamics between the qubits for the initial state $\ket{\Psi_{+}}$ will be the same as entanglement dynamics between the qubits for the initial state $\ket{\Phi_{+}}$. One should note that this result is valid for any initial state of the field.  Because of this, we explicitly study only the states for which the initial state of the qubits is $\ket{\Psi_{+}}$. 

When the qubits start in the state $\ket{\Psi_{+}}$, it turns out that the expression for concurrence \cite{Wootters}, a measure that  quantifies entanglement between two qubits, has an analytic general form that is valid for all the initial states of the oscillator. We first derive that general expression and then study explicitly the cases for which the initial state of the oscillator is either a thermal state, a coherent state or a number state.

For the state $\ket{\Psi_{+}}$, the initial density matrix for the qubits, $\hat{Q}$, takes the form:
\beq
\hat{Q} =\frac{1}{2}
\begin{pmatrix}
1 & 0 & 0 & -1  \\
 0 & 0 & 0 & 0  \\
 0 & 0 & 0 & 0  \\
 -1 & 0 & 0 & 1  \\
\end{pmatrix},
\eeq
where the rows and columns are defined in the basis: $\ket{1,1}$, $\ket{1,0}$, $\ket{0,0}$ and $\ket{1,-1}$. 
We denote the initial state of the oscillator by $\hat{F}$, which is characterized by the Glauber-Sudarshan function  \textit{P}.
As time evolves, by using equation (\ref{e.qubit_dens_mat}), we calculate the reduced density matrix for the qubits in the $\ket{j,m}$ basis and get:
\beq
\hat{\rho}_q(t) =\frac{1}{2}
\begin{pmatrix}
 1 & 0 & 0 & -I(t)  \\
 0 & 0 & 0 & 0  \\
 0 & 0 & 0 & 0  \\
 -I(t)^* & 0 & 0 & 1  \\
\end{pmatrix},
\eeq
where 
\beqa\label{e.integral}
I(t)&=&\int \mathrm{d^2}\alpha P(\alpha)\exp{(-8i\beta f(\omega t,\alpha))}\nonumber\\
&\times&\exp{\left(-32\beta^2\sin^{2}\left(\omega t/2\right)\right)}.
\eeqa

In order to calculate the concurrence in the usual form between the two qubits, we need to write $\hat{\rho}_q(t)$ in the $\ket{e,e}, \ket{e,g}, \ket{g,e}$ and $\ket{g,g}$ basis:
\beq\label{e.Den-mat}
\hat{\rho}_q(t)=\frac{1}{4}
\begin{pmatrix} 1-u(t) & i v(t) & i v(t) & 1-u(t) \\ 
                 -i v(t) & 1+u(t) & 1+u(t) & -i v(t) \\
                 -i v(t) & 1+u(t) & 1+u(t) & -i v(t) \\
                 1-u(t) & i v(t) &  i v(t) & 1-u(t) \\
\end{pmatrix}.
\eeq
The terms $u(t)$ and $v(t)$ are the real and imaginary parts of $I(t)$ respectively. 
For this density matrix, the concurrence is given by:
\beqa\label{e.Con_Gen}
C(t)&=&\left|I(t)\right|,\nonumber\\
&=&\left|\int \mathrm{d^2}\alpha P(\alpha)\exp{(-8i\beta f(\omega t,\alpha))}\right|\nonumber\\
&&\times\exp{\left(-32\beta^2\sin^{2}\left(\omega t/2\right)\right)},\nonumber\\
&=&\left|\tilde{P}\left(8\beta\sin(\omega t),8\beta(\cos(\omega t)-1)\right)\right|\nonumber\\
&&\times\exp{\left(-32\beta^2\sin^{2}\left(\omega t/2\right)\right)}.
\eeqa
The function $\tilde{P}$, defined in equation (\ref{e.Con_Gen}), is the Fourier transform of P:
\beq
\tilde{P}(k_{1},k_{2})=\int \mathrm{d}x\mathrm{d}y P(x,y)\exp{(-i(k_{1}x+k_{2}y))},
\eeq 
with $x$ and $y$ being respectively the real and imaginary parts of $\alpha$ in the integral (\ref{e.integral}). 
From equation (\ref{e.Con_Gen}), we see that knowing $P(\alpha)$ or its Fourier transform, $\tilde{P}$, that characterizes the initial state of the oscillator, we can follow the evolution of entanglement between the qubits.

\vspace{0.3cm}

\textbf{Qubits initially in state $\ket{\Psi_{+}}$ and oscillator in the thermal state}:   
The dynamics of a system interacting with a thermal reservoir is generally associated with decoherence of the system. However, following the work of Bose et al. \cite{Bose-etal}, it was shown in \cite{Kim-etal} that entanglement (which is a manifestation of coherent superposition of states of a multi-partite system), can be created between the two qubits of the Tavis-Cummings model with the oscillator initially prepared in the thermal state. Environment induced entanglement is an area of active research \cite{Oh-Kim, Env-Ind-Ent}. 

Entanglement dynamics of the two-qubits Tavis-Cummings model with the oscillator initially in a thermal state is investigated in \cite{Aguiar, Zhou, Roa-etal, Jin-etal}. In particular, Roa et al. \cite{Roa-etal}, studied the entanglement dynamics between the qubits in the dispersive coupling regime within the rotating wave approximation, where the qubits were taken to be far off resonance from the oscillator and the coupling between the qubits and the oscillator taken to be weak.

To study the entanglement dynamics between the qubits in the degenerate regime, beyond the RWA and for arbitrary coupling strengths, for the initial state of the qubits being $\ket{\Psi_{+}}$ and that of the oscillator being a thermal state using equation (\ref{e.Con_Gen}), we note that for a thermal state $P(\alpha)$ is given by \cite{Scully-Zubairy}:
\beq\label{e.P_Th}
P_{th}(\alpha)=\frac{1}{\pi\left\langle n\right\rangle}\exp{(-|\alpha|^{2}/\left\langle n\right\rangle)}.
\eeq
In the above equation, $\left\langle n\right\rangle$ is the average number of excitations for a given temperature which characterizes the thermal state. We note that as $P(\alpha)$ is a gaussian of width $\sqrt{\left\langle n\right\rangle}$, its Fourier transform, $\tilde{P}$, is also a gaussian of width $1/\sqrt{\left\langle n\right\rangle}$.

\begin{figure}
	\centering
	\subfigure{\includegraphics[width=9cm]{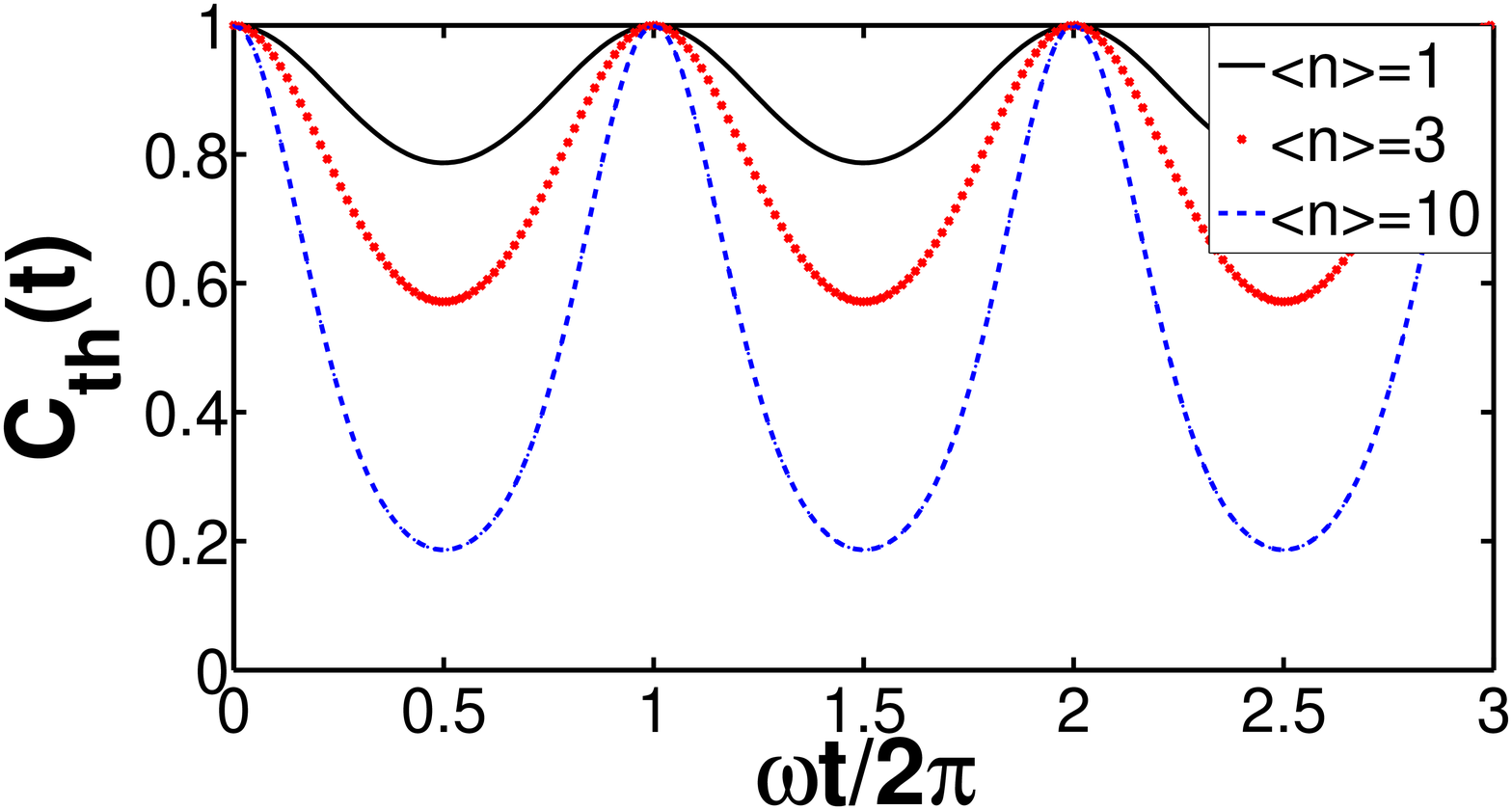}}
	\subfigure{\includegraphics[width=9cm]{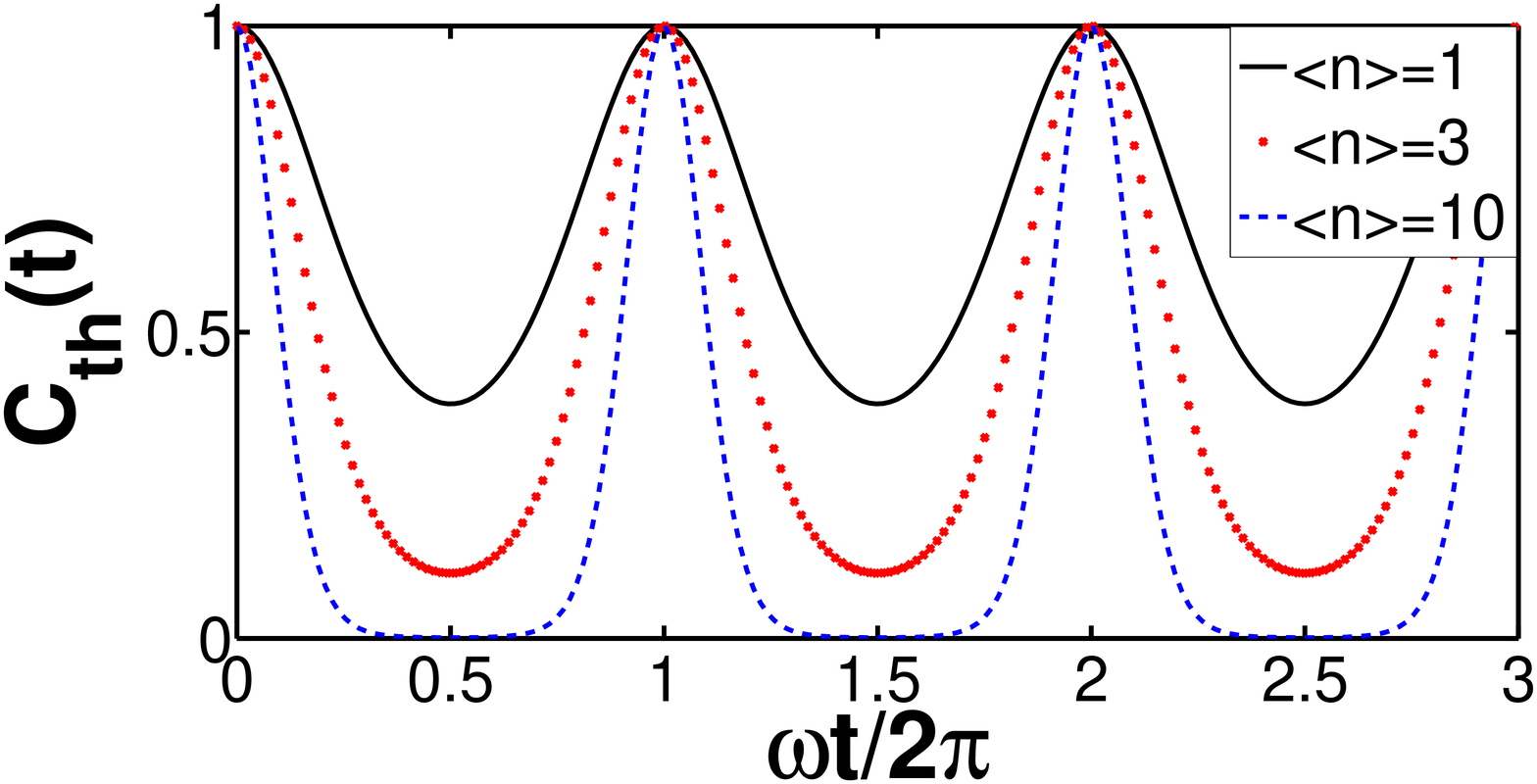}}
	\caption{Time dependence of entanglement between the two qubits in the degenerate regime given by $C_{th}(t)$ for various value of $\left\langle n\right\rangle$. Top: $\beta=0.05$ and bottom: $\beta=0.1$.}
	\label{f.Degen_Th}
\end{figure}

Using $P_{th}(\alpha)$ in equation (\ref{e.Con_Gen}), we get: 
\beq\label{e.C_th}
C_{th}(t)=\exp{\left(-32\beta^2\sin^{2}{(\omega t/2)}(1+2\left\langle n\right\rangle)\right)}.
\eeq
In Fig. \ref{f.Degen_Th}, $C_{th}(t)$ is plotted for various values of $\left\langle n\right\rangle$ and $\beta=0.05$ and $\beta=0.1$. We notice that entanglement between the qubits oscillates periodically with time period $2\pi/\omega$. Within each time period, the entanglement between the qubits decreases more rapidly from its maximum value as one increases $\left\langle n\right\rangle$ or $\beta$. The minimum entanglement between the qubits, $\exp{\left(-32\beta^2(1+2\left\langle n\right\rangle)\right)}$, reduces as  $\left\langle n\right\rangle$ or $\beta$ is increased. These are consequences of $\tilde{P}$ becoming narrower as $\left\langle n\right\rangle$ or $\beta$ increases. 

\vspace{0.3cm}

\textbf{Qubits initially in state $\ket{\Psi_{+}}$ and oscillator in a coherent state}: 
The dynamics of a single qubit interacting with a coherent state of the oscillator has been studied extensively in the literature and has revealed many important phenomena that are unique features of quantum mechanics such as collapse and revival of atomic coherence \cite{Collapse-Revival}, entanglement and disentanglement of the qubit with the oscillator and in process preparation of the Schr\"odinger Cat state \cite{Ent-Sch-Cat}, etc. 
Within the RWA, the collapse and revival of a two-qubit TCM was studied in \cite{Deng85}, and the entanglement dynamics for various different initial states of the oscillator, including coherent state, was done in \cite{Tessier_etal}.

\begin{figure}
\includegraphics[width= 9cm]{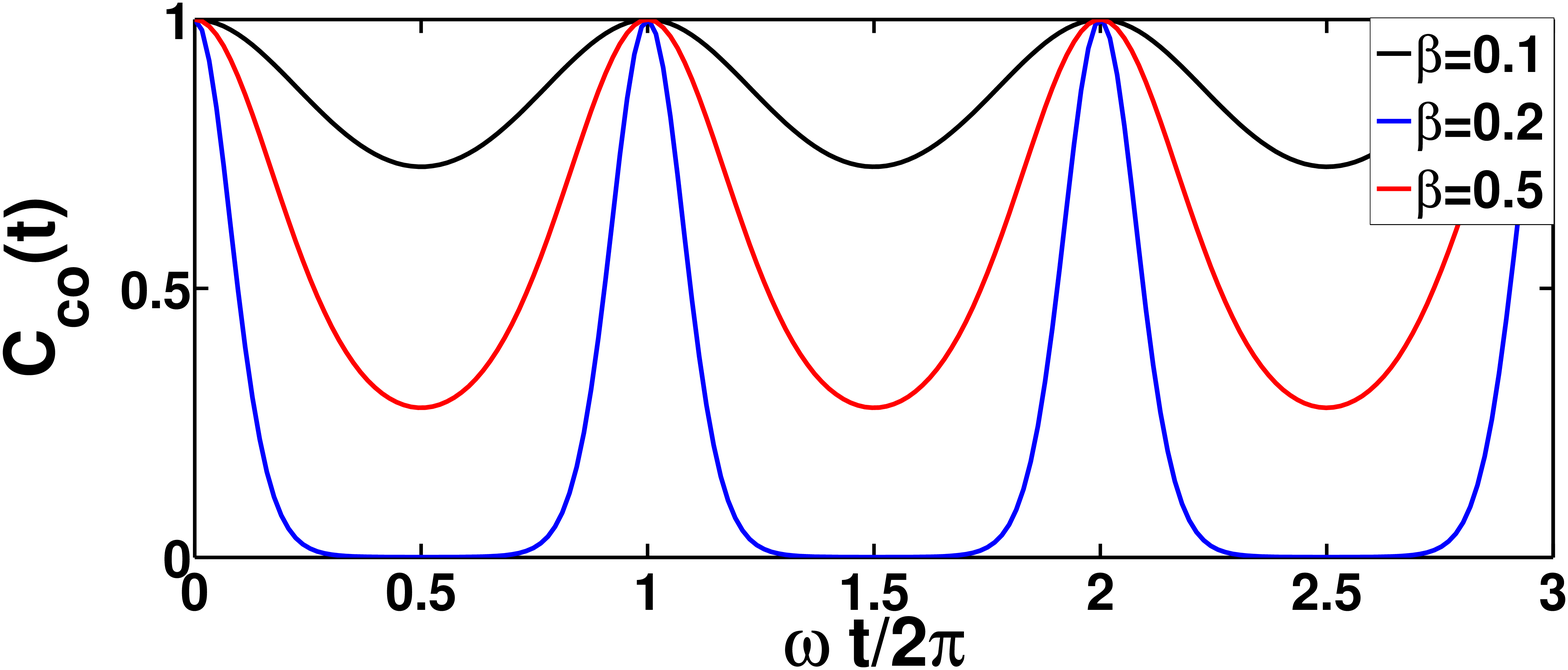}
\caption{Time dependence of entanglement between the two qubits in the degenerate regime given by $C_{co}(t)$ for various value of $\beta$.} \label{f.Degen_Coh}
\end{figure}

For the oscillator initially in the coherent state $\ket{\alpha_{0}}$, $P(\alpha)$ is given by \cite{Scully-Zubairy}: 
\beq\label{e.P_Co}
P_{coh}(\alpha)=\delta^{2}(\alpha-\alpha_{0}).
\eeq
With $P_{coh}(\alpha)$, $I(t)$ in equation (\ref{e.integral}) is evaluated to be:
\beqa\label{e.integral_Co}
I_{coh}(t)&=&
\exp{\left(-8i\beta f(\omega t,\alpha_{0})\right)}\nonumber\\
&&\times\exp{\left(-32\beta^2\sin^{2}{(\omega t/2)}\right)}.
\eeqa
From equations (\ref{e.Con_Gen}) and (\ref{e.integral_Co}), and the fact that $f(\omega t,\alpha_0)$ is real, the concurrence for the coherent state case is found to be:
\beq\label{e.C_co}
C_{coh}(t)=\exp{\left(-32\beta^2\sin^{2}{(\omega t/2)}\right)}.
\eeq
We see from equation (\ref{e.C_co}) that the entanglement between the qubits does not depend on the  oscillator strength, $\alpha_0$, and only depends upon $\beta$ and $\omega$. The fact that $C_{coh}(t)$ does not depend upon $\alpha_{0}$ is a consequence of $\tilde{P}$ being just a phase as $P_{coh}(\alpha)$ is a delta function. In Fig. \ref{f.Degen_Coh}, we plot $C_{coh}(t)$ for various values of $\beta$. We see from Fig. \ref{f.Degen_Coh} that $C_{coh}(t)$ is periodic with time period $2\pi/\omega$. As in the thermal state case, the entanglement between the qubits decreases more rapidly from its maximum value as one increases $\beta$. 

\vspace{0.3cm}

\textbf{Qubits initially in state $\ket{\Psi_{+}}$ and oscillator in a number state}:   
To study the entanglement dynamics between the qubits in the degenerate regime when the oscillator starts in the Fock state, we note that $P({\alpha})$ for a Fock state $\ket{N}$ is given by \cite{Scully-Zubairy}:
\beq\label{e.P_Num}
P_{N}(\alpha)=\frac{\exp{(\alpha\alpha^{*})}}{N!}\left(\frac{\partial^{2N}}{\partial\alpha^{N}\partial\alpha^{*N}}\delta^{2}(\alpha)\right).
\eeq
With $P_{N}(\alpha)$ in equation (\ref{e.integral}), $I(t)$ becomes (see Appendix B): 
\beqa\label{e.integral_Num}
I_{N}(t)&=&
\sum_{m=0}^{N}\binom{N}{m}\frac{1}{(N-m)!}\left(8i\beta\sin{(\omega t/2)}\right)^{2(N-m)}\nonumber\\
&&\times\exp{\left(-32\beta^2\sin^{2}(\omega t/2)\right)},\nonumber\\
&=&L_{N}\left((8\beta\sin{(\omega t/2)})^2\right)\nonumber\\
&&\times\exp{\left(-32\beta^2\sin^{2}(\omega t/2)\right)},
\eeqa
where $L_{N}(x)$ is a Laguerre polynomial.
Using equations (\ref{e.integral_Num}) and (\ref{e.Con_Gen}), we calculate the concurrence for the number state case:
\beqa\label{e.Con_N}
C_{N}(t)&=&\left|L_{N}\left((8\beta\sin{(\omega t/2)})^2\right)\right|\nonumber\\
&&\times\exp{\left(-32\beta^2\sin^{2}(\omega t/2)\right)}.
\eeqa

\begin{figure}[htb]
	\centering
	\subfigure{\includegraphics[width=9cm]{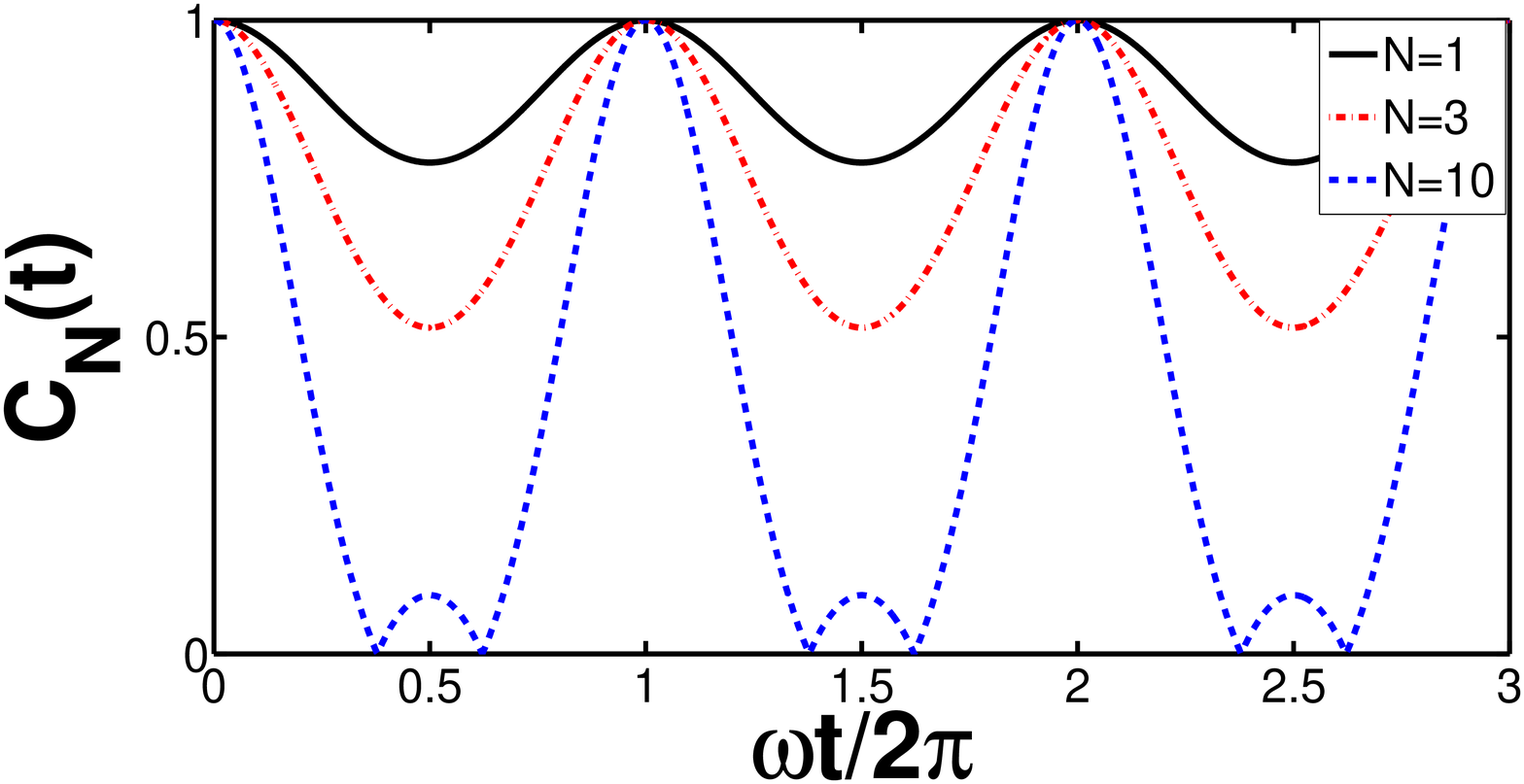}}
	\subfigure{\includegraphics[width=9cm]{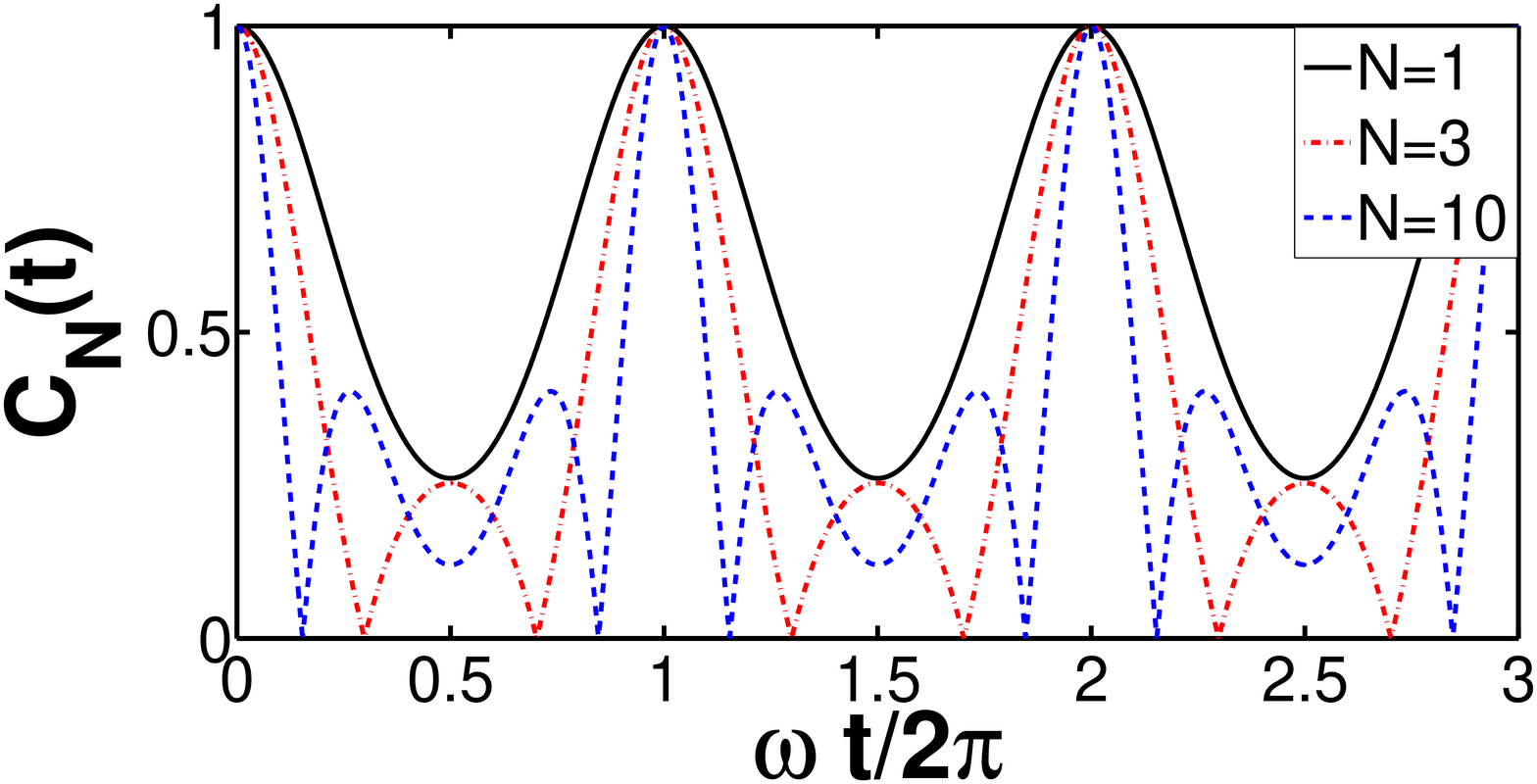}}
	\caption{Time dependence of entanglement between the two qubits in the degenerate regime given by $C_{N}(t)$ for various value of $N$. Top: $\beta=0.05$ and bottom: $\beta=0.1$.}
	\label{f.Degen_Num}
\end{figure}

The concurrence, $C_{N}(t)$, is plotted in Fig. \ref{f.Degen_Num} for various values of $N$ and $\beta=0.05$ and $\beta=0.1$. We see from the figure that $C_{N}(t)$ is periodic with time period $2\pi/\omega$. Within each time period, for certain values of $N$ and $\beta$, the qubits momentarily become disentangled before getting entangled again. As $N$ or $\beta$ is increases, entanglement between the qubits gets sharply peaked at times that are integer multiples of $2\pi/\omega$.     

\vspace{0.3cm}

\begin{figure}[htb]
	\centering
	\subfigure{\includegraphics[width=7cm]{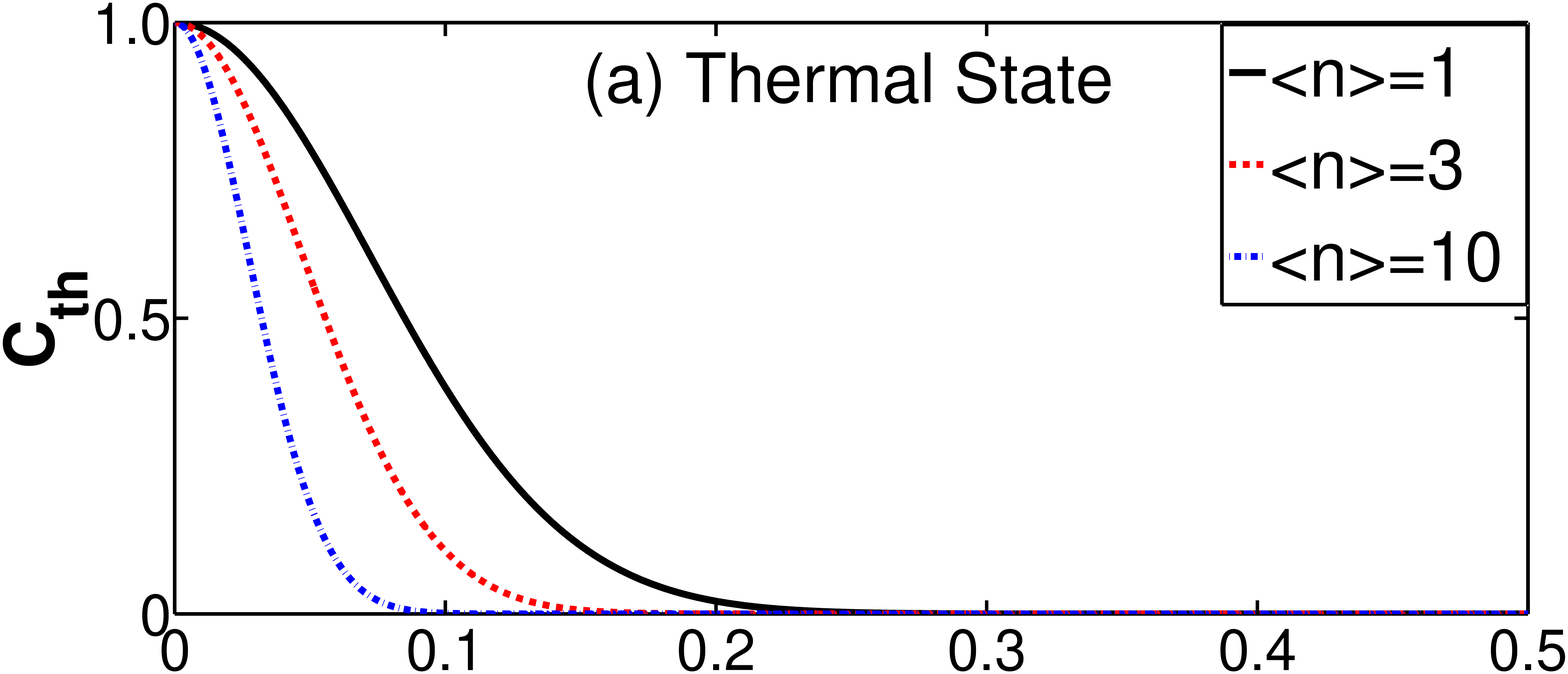}}
	\subfigure{\includegraphics[width=7cm]{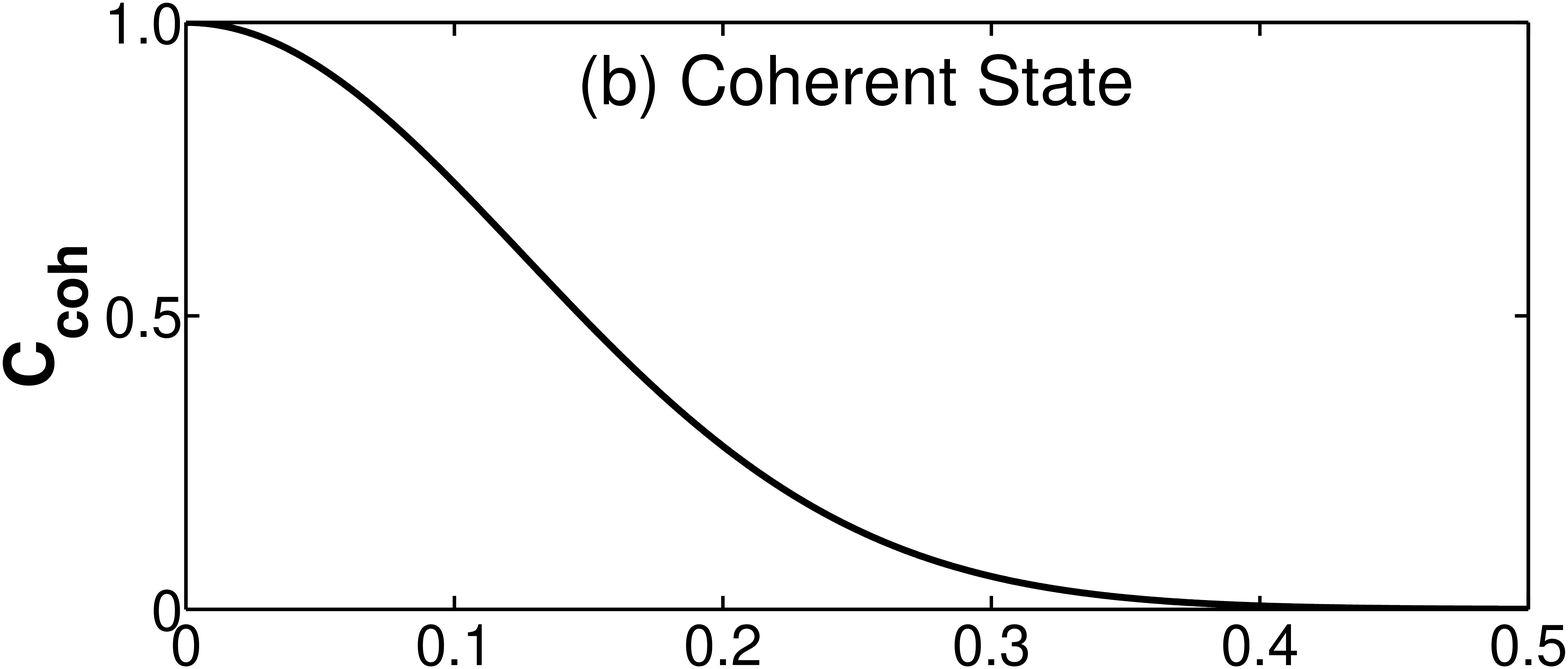}}
	\subfigure{\includegraphics[width=7cm]{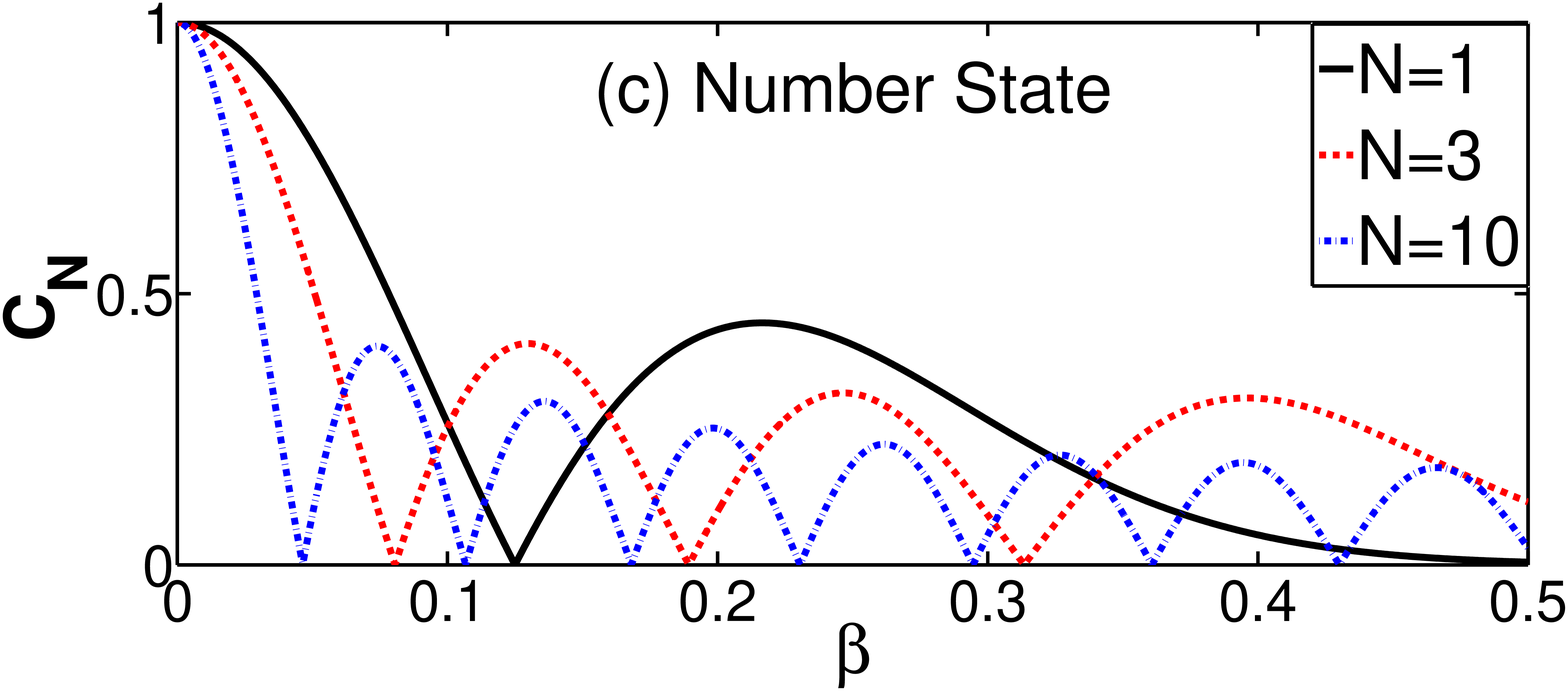}}
	\caption{Entanglement between the qubits as a function of $\beta$ for (a) thermal state, (b) coherent state, and (c) number state evaluated at $\omega t/2=\pi/2$.}
	\label{f.Depend_beta}
\end{figure}

\textbf{Dependence of entanglement on coupling strength}:
As discussed in section \ref{s.model}, for the RWA to be valid, one has to ensure that $\lambda\ll\omega,\omega_0$. So, given $\omega$ and $\omega_0$, there is an upper limit of $\lambda$ for which RWA is valid. In contrast to the RWA case, the results in this report are correct for all values of $\lambda$. 
We see from equations (\ref{e.C_th}), (\ref{e.C_co}) and (\ref{e.Con_N}) that the entanglement depends on the coupling strength and the oscillator frequency through the parameter $\beta^2\sin^{2}(\omega t/2)$. 
This parameter is periodic with time period $T=2\pi/\omega$. At times that are integer multiples of $T$, the qubits are maximally entangled for all coupling strengths. This can be seen from Figs. \ref{f.Degen_Th}, \ref{f.Degen_Coh} and \ref{f.Degen_Num}. To understand the role of the coupling strength, $\lambda$, in the entanglement between the qubits, we plot in Fig. \ref{f.Depend_beta} the concurrence for the thermal state, the coherent state and the number state as a function of the dimension-less coupling strength $\beta$ evaluated at half the time period, i.e. $t=\pi/\omega$. We have: 
\beqa
C_{th}(\pi/\omega)&=&\exp{\left(-32\beta^2(1+2\left\langle n\right\rangle)\right)},\\
C_{coh}(\pi/\omega)&=&\exp{\left(-32\beta^2\right)},\\
C_{N}(\pi/\omega)&=&\left|L_{N}\left(64\beta^2\right)\right|\exp{\left(-32\beta^2\right)}.
\eeqa

We see from the above set of equations that for all the three initial states of the oscillator, concurrence decreases exponentially as $\beta$ increases. Equation ($37$), for various values of $\left\langle n \right\rangle$, is plotted in Fig. \ref{f.Depend_beta}(a) from which we see that the concurrence decreases faster as the mean thermal excitation number $\left\langle n \right\rangle$ is increased. For the coherent state, equation ($38$) is plotted in Fig. \ref{f.Depend_beta}(b) from which the exponential decrease of concurrence is evident. In Fig. \ref{f.Depend_beta}(c), $C_{N}(\pi/\omega)$ is plotted for various values of $N$ where we see that the exponential decrease of concurrence is modulated by the Laguerre polynomial.    

\section{Classical-Quantum Comparison}\label{s.comparison}

It is well established within the rotating wave approximation that the predictions about the dynamical properties of the qubits interacting with a classical harmonic oscillator are qualitatively different  from the predictions when the oscillator is treated quantum mechanically. Contrary to the classical case involving a prescribed driving field, the quantum mechanical treatment for the corresponding coherent field predicts the qubit population inversion to show collapse and revival \cite{Collapse-Revival}. These arise from the infinitely many, but discrete, Rabi frequencies for qubit evolution associated with the infinite set of Fock states defining a coherent state. The quantum mechanical predictions have been confirmed experimentally \cite{Brune-etal}. In view of this, it is a question whether there are qualitative differences between the predictions that can be made on the basis of the non-RWA calculations in Sec. IV and a set of corresponding non-RWA predictions to be made when the oscillator is treated as classical, while the qubits remain quantum mechanical.

A mixed quantum-classical theory of this kind is usually called semi-classical and the oscillator is taken as an external influence. Here we want something a bit different because in Sec. IV the oscillator has not been treated as an external influence, but rather as open to back reaction from the qubits. Jaynes first addressed this issue of back reaction in the cavity QED context with a so-called ``neo-classical" theory \cite{Jaynes, JC} in which the oscillator (cavity electric field in his case) was driven by the qubit via its quantum average values. In this way, although influenced by the quantized qubits, the oscillator remains unquantized for all time and takes only numerical, not operator, values.

We will adopt the Jaynes prescription in order to include back reaction. The simplest approach is to derive equations of motion for the various dynamical quantities using the original Hamiltonian (\ref{e.H_o}). For the oscillator, instead of working with non-hermitian operators $\hat{a}$ and $\hat{a}^{\dagger}$, we work with the real dynamical quantities, momentum $\hat{p}$ and co-ordinate $\hat{q}$ defined in Appendix C. We then replace operators by numerical variables, which we indicate simply by removing the carets. These variables evolve from classically specifiable initial values $p(0)$ and $q(0)$, as driven by expectation values of the qubit $\hat{\sigma}_{x}$ variables:
\beqa\label{e_evol}
\dot{q}&=&p+\hbar\lambda\sqrt{2/\hbar\omega}\langle{\sigma}_{x}^{(1)}+\hat{\sigma}_{x}^{(2)}\rangle,\label{e.q_evol}\\
\dot{p}&=&-\omega^{2}q,\label{e.p_evol}\\
\dot{\hat{\sigma}}_{x}^{(i)}&=&0,\\
\dot{\hat{\sigma}}_{y}^{(i)}&=&-2\lambda\sqrt{2/\hbar\omega}\mbox{ }p\hat{\sigma}_{z}^{(i)},\label{e.s_y_evol}\\
\dot{\hat{\sigma}}_{z}^{(i)}&=&2\lambda\sqrt{2/\hbar\omega}\mbox{ }p\hat{\sigma}_{y}^{(i)}.\label{e.s_z_evol}
\eeqa

As it happens, the states we have worked with have zero expectation of the $\hat{\sigma}_{x}$ operators, which don't evolve themselves, so the oscillator evolves completely independently. For the qubits starting in the state $\ket{\Psi_{+}}$, there are only four non-zero density matrix elements: $\matel{1,\pm1}{\hat{\rho}_{q}}{1,\pm1}$ and $\matel{1,\pm1}{\hat{\rho}_{q}}{1,\mp1}$. The diagonal density matrix elements do not change in time because $\dot{\hat{\sigma}}_{x}^{(1)}+\dot{\hat{\sigma}}_{x}^{(2)}=0$. To find the evolution of the off-diagonal elements, we note that:
\beqa\label{e.mat_el_SS}
\matel{1,1}{\hat{\rho}_{q}(t)}{1,-1}&=&Tr\{\hat{\rho}_{q}(t)\ket{1,-1}\bra{1,1}\}\nonumber\\
&=&Tr\{\hat{\rho}_{q}(0)\hat{S}^{(1)}_{+}(t)\hat{S}^{(2)}_{+}(t)\},\nonumber\\
&=&\left\langle\hat{S}^{(1)}_{+}(t)\hat{S}^{(2)}_{+}(t)\right\rangle,
\eeqa
where 
\beq
\hat{S}^{(i)}_{+}(t)=\frac{1}{2}(\hat{\sigma}_{z}^{(i)}(t)+i\hat{\sigma}_{y}^{(i)}(t)).
\eeq
From equations (\ref{e.s_y_evol}) and (\ref{e.s_z_evol}), we get the following equation for the required expectation values
\beq\label{e.S1S2_evol_cl}
\frac{d}{dt}\la\hat{S}^{(1)}_{+}(t)\hat{S}^{(2)}_{+}(t)\ra=-4i\lambda\sqrt{\frac{2}{\hbar\omega}}\mbox{ }p\la\hat{S}^{(1)}_{+}(t)\hat{S}^{(2)}_{+}(t)\ra.
\eeq
Solving equations (\ref{e.q_evol}), (\ref{e.p_evol}) and (\ref{e.S1S2_evol_cl}) we get
\beqa\label{e.S1S2_evol_final}
\la\hat{S}^{(1)}_{+}(t)\hat{S}^{(2)}_{+}(t)\ra&=&\la\hat{S}^{(1)}_{+}(0)\hat{S}^{(2)}_{+}(0)\ra\nonumber\\
&\times&\exp{\left(8i\beta\sqrt{2\omega/\hbar}\mbox{ }q(0)\sin^{2}(\omega t/2)\right)}\nonumber\\
&\times&\exp{\left(-4i\beta\sqrt{2/\hbar\omega}\mbox{ }p(0)\sin(\omega t)\right)},
\eeqa
where $q(0)$ and $p(0)$ are the initial position and momentum of the oscillator respectively. 
For the initial state of the qubits being $\ket{\Psi_{+}}$, $\la\hat{S}^{(1)}_{+}(0)\hat{S}^{(2)}_{+}(0)\ra=-1/2$.
The entanglement between the qubits is then given by (equations (\ref{e.Con_Gen}) and (\ref{e.mat_el_SS})):
\beq
C(t)=2|\la\hat{S}^{(1)}_{+}(t)\hat{S}^{(2)}_{+}(t)\ra|.
\eeq
Since the time dependent terms in equation (\ref{e.S1S2_evol_final}) enter only as phases, the entanglement between the qubits do not change and the qubits remain maximally entangled.  

A crucial point used in deriving equation (\ref{e.S1S2_evol_final}) is the fact that the oscillator is assumed to have well defined initial position and momentum. Because of the uncertainty principle, within quantum mechanics, no state can be a simultaneous eigenstate of position and momentum. 
In view of this, it is a question whether with a classical description of the oscillator one can or cannot get modulations in entanglement of the qubits if the initial position and momentum of the oscillator are not well defined. 
We now show that by averaging equation (\ref{e.S1S2_evol_final}) over appropriate joint probability distributions of $q(0)$ and $p(0)$, say $P(q_{0},p_{0})$, we get expressions for concurrence that match with the quantum mechanically derived results for thermal states (equation (\ref{e.C_th})) and the coherent states (equation (\ref{e.C_co})). 

The classical probability distribution $P(q_{0},p_{0})$ of a harmonic oscillator in thermal equilibrium is gaussian. For an appropriate probability distribution of the oscillator in a coherent state, $\ket{\overline{\alpha}_{0}}$, we use the Wigner distribution of the coherent state \cite{Scully-Zubairy} which is also a gaussian. So, for the purpose of thermal states and coherent states, we take $P(q_{0},p_{0})$ to be of the general form:
\beq
P(q_{0},p_{0})=\frac{1}{\pi\Delta q\Delta p}\exp{\left(-\frac{(q_{0}-\bar{q}_{0})^{2}}{(\Delta q)^{2}}-\frac{(p_{0}-\bar{p}_{0})^{2}}{(\Delta p)^{2}}\right)},
\eeq 
where $\Delta q$ and $\Delta p$ are the initial uncertainties in position and momentum, centered at $\bar{q}_{0}$ and $\bar{p}_{0}$ respectively. 
Taking the absolute value after averaging equation (\ref{e.S1S2_evol_final}) over $P(q_{0},p_{0})$, we get
\beqa\label{e.C_cl}
\bar{C}(t)&=&2\left|\int \mathrm{d\textit{q}_{0}d\textit{p}_{0}}P(q_{0},p_{0})\la\hat{S}^{(1)}_{+}(t)\hat{S}^{(2)}_{+}(t)\ra\right|,\nonumber\\
&=&\exp{\left(-32\beta^2\sin^{4}\left(\omega t/2\right)\frac{\omega (\Delta q)^{2}}{\hbar}\right)}\nonumber\\
&&\times\exp{\left(-8\beta^2\sin^{2}\left(\omega t\right)\frac{(\Delta p)^{2}}{\hbar\omega}\right)}.
\eeqa 
From the above equation, we see that the concurrence does not depend on the average values $\bar{q}_{0}$ and $\bar{p}_{0}$ and only depends on the uncertainties, $\Delta q$ and $\Delta p$. If both the uncertainties, $\Delta q$ and $\Delta p$, are not simultaneously zero, the concurrence does change in time.

For a thermal state characterized by temperature $T$, $\Delta q_{th}=\sqrt{2KT/\omega^2}$ and $\Delta p_{th}=\sqrt{2KT}$ where $K$ is the Boltzmann constant. Using this in equation (\ref{e.C_cl}), we get:
\beq\label{e.C_cl_th}
\bar{C}_{th}(t)=\exp{\left(-64\beta^2\sin^{2}\left(\omega t/2\right)\frac{KT}{\hbar\omega}\right)}.
\eeq 
In order to compare equation (\ref{e.C_cl_th}) with the quantum result (equation(\ref{e.C_th})), we note that in the high temperature limit, i.e. when $KT\gg\hbar\omega$, the average excitation number $\la n\ra=KT/\hbar\omega$. Now, replacing the factor $(1+2\la n\ra)$ in the exponential of equation (\ref{e.C_th}) by $2KT/\hbar\omega$, we get exactly the same  expression for concurrence as derived with the classical description of the oscillator, equation (\ref{e.C_cl_th}). 

A coherent state is a minimum uncertainty state for which $\Delta q_{coh}=\sqrt{\hbar/\omega}$ and $\Delta p_{coh}=\sqrt{\hbar\omega}$. Putting this in equation (\ref{e.C_cl}), we get:
\beq\label{e.C_cl_co}
\bar{C}_{coh}(t)=\exp{\left(-32\beta^2\sin^{2}\left(\omega t/2\right)\right)},
\eeq 
which exactly matches with the corresponding quantum result, equation (\ref{e.C_co}).  

The above results show that the modulations in the entanglement between the qubits is a consequence of initial uncertainty in $q_{0}$ and $p_{0}$, and not an intrisically quantum property. However, the uncertainty in the simultaneous assignment of values to canonically conjugate variables is inherent in quantum mechanics and cannot be avoided. This fundamental uncertainty is reflected in the general equation for concurrence that is valid for all quantum states, equation (\ref{e.Con_Gen}), from which we see that for all states there is a time dependent factor, $\exp{\left(-32\beta^2\sin^{2}\left(\omega t/2\right)\right)}$, which corresponds to the modulation of entanglement for the minimum uncertainty state, i.e. the coherent state.




\section{Conclusion}\label{s.conclusion}
In this report, we studied the dynamics of the two-qubit Tavis-Cummings model in the degenerate regime beyond the RWA, obtaining exact analytic expressions for state properties of interest, including qubit entanglement. The general form of the 
time evolved density matrix of the composite system for any initial product state of the qubits and the oscillator was derived. The qubits initially prepared in the $\ket{\Psi_{-}}$ and $\ket{\Phi_{-}}$ states were shown to remain maximally entangled throughout the dynamical evolution. The dynamics of entanglement between the qubits initially prepared in the $\ket{\Psi_{+}}$ state was shown to be the same as that for the qubits initially prepared in the state $\ket{\Phi_{+}}$ for any initial state of the oscillator. A general expression for concurrence, quantifying the entanglement between the qubits initially prepared in the state $\ket{\Psi_{+}}$ and the oscillator in any state characterized by the Glauber-Sudarshan function $P(\alpha)$ was derived. 

When the initial state of the oscillator is either the thermal state, coherent state or number state, the entanglement dynamics between the qubits was shown to be periodic with time period $2\pi/\omega$. This periodic behavior is a consequence of the fact that our system consists of displaced harmonic oscillators with frequency $\omega$. A general feature of entanglement for all initial states of the oscillator was a more sharply peaked entanglement as the dimension-less coupling strength $\beta$ increased. This feature can be understood as a manifestation of the common factor $\exp{\left(-32\beta^2\sin^{2}\left(\omega t/2\right)\right)}$ in equation (\ref{e.Con_Gen}), which is present for all initial states of the oscillator. 

When the initial state of the oscillator is a coherent state, it was shown that the entanglement between the qubits does not depend on the oscillator strength $\alpha_{0}$. The periodicity of entanglement between the qubits with a period independent of the strength of the coherent state was reported in \cite{Roa-etal} where the dynamics between the qubits and the oscillator was studied in the dispersive regime within the RWA. Although in 
\cite{Roa-etal} the periodicity was found not to depend on the strength of the coherent state, the amplitude of oscillations of entanglement depended on $\alpha_{0}$ in contrast to the results found here in the degenerate regime.  

In section \ref{s.comparison}, we studied the dynamics of the state of the qubits by treating the oscillator classically. The back action of the qubits on the oscillator evolution equations was included in the spirit of Jaynes' original neoclassical treatment \cite{Jaynes}. It was shown that temporal entanglement modulation is also present classically, but only if there is uncertainty in the initial position and/or momentum of the oscillator. 

In all the calculations in the current paper, the qubits were assumed to be perfectly degenerate. This is obviously an idealization. A more realistic scenario would have quasi-degenerate qubits with $\omega_{0} \neq 0$ but still satisfying the condition $\omega_{0} \ll \omega,\lambda$. The existence of comparable analytic expressions for the collective properties of such qubits with non-zero $\omega_{0}$ is an important question for future work.




\section*{Acknowledgement}\label{Acknowl}

Partial financial support was received from DARPA HR0011-09-1-0008, ARO W911NF-09-1-0385, and NSF PHY-0855701.
\bigskip


\noindent APPENDIX A: EVOLUTION TERM\label{a.evolution}

\beqa\label{e.evolution}
&& e^{-i\hat{H}_0 t/\hbar} \ket{j,m}\ket{\alpha} = \nonumber \\
&=& e^{-i(\omega \hat{a}^{\dagger}\hat{a}
+ \lambda(\hat{a}+\hat{a}^{\dagger}) (\hat{\sigma}_{x}^{1}+\hat{\sigma}_{x}^{2}))t}\ket{j,m}\ket{\alpha},\nonumber\\
&=& e^{-i(\omega \hat{a}^{\dagger}\hat{a}
+ \lambda(\hat{a}+\hat{a}^{\dagger})2m)t}\ket{j,m}\ket{\alpha},\nonumber\\
&=&\ket{j,m}\hat{D}(-2\beta_m)
e^{-i\omega t \hat{a}^{\dagger}\hat{a}}\hat{D}(2\beta_m)\ket{\alpha}e^{4i\beta_m^2 \omega t},\nonumber\\
&=&\ket{j,m}\ket{(\alpha+2\beta_m)e^{-i\omega t}-2\beta_m}\nonumber\\
&\times&\exp{(-2i\beta_m\sin{\frac{\omega t}{2}}(\alpha^*e^{i\omega t/2}+\alpha e^{-i\omega t/2}))}\nonumber\\
&\times&\exp{(4i\beta_m^2(\omega t-\sin{\omega t}))}.
\eeqa\\

\noindent APPENDIX B: CALCULATION OF $I_{N}(t)$\label{a.I_N}

We define
\beq
\tilde{I}_{N}(t)=I_{N}(t)\exp{\left(-32\beta^2\sin^{2}\left(\omega t/2\right)\right)},
\eeq
and evaluate $\tilde{I}_{N}(t)$ for $P_{N}(\alpha)$ given by equation (\ref{e.P_Num}):
\beqa
\tilde{I}_{N}(t)&=&\int \mathrm{d^2}\alpha P_{N}(\alpha) \nonumber \\
&\times& \exp{(-8i\beta\sin{\frac{\omega t}{2}}(\alpha^*e^{i\omega t/2}+\alpha e^{-i\omega t/2}))} \nonumber\\
&=&\frac{1}{N!}\int \mathrm{d^2}\alpha \exp{(|\alpha|^2)}\left(\frac{\partial^{2N}}{\partial\alpha^{N}\partial\alpha^{*N}}\delta^{2}(\alpha)\right)\nonumber\\
&\times&\exp{(-8i\beta\sin{\frac{\omega t}{2}}(\alpha^*e^{i\omega t/2}+\alpha e^{-i\omega t/2}))} \nonumber\\
&=&\frac{1}{N!}\int \mathrm{d^2}\alpha \exp{(|\alpha|^2)}\left(\frac{\partial^{2N}}{\partial\alpha^{N}\partial\alpha^{*N}}\delta^{2}(\alpha)\right)\nonumber\\
&\times&\exp{(A(\alpha^*a+\alpha b))}.
\eeqa
where
\beqa
A&=&-8i\beta\sin{\frac{\omega t}{2}},\nonumber\\
a &=& e^{i\omega t/2},\quad {\rm and}\quad b = e^{-i\omega t/2}.
\eeqa
Using the identity
\beqa
&&\int \mathrm{d^2}\alpha \left(\frac{\partial^{2N}}{\partial\alpha^{N}\partial\alpha^{*N}}\delta^{2}(\alpha)\right)f(\alpha,\alpha^{*})\nonumber\\
&=&\left(\frac{\partial^{2N}}{\partial\alpha^{N}\partial\alpha^{*N}}f(\alpha,\alpha^{*})\right)|
_{\alpha=\alpha^{*}=0},
\eeqa
we get
\begin{align}
\tilde{I}_{N}(t)&=\frac{1}{N!}\frac{\partial^{2N}}{\partial\alpha^{N}\partial\alpha^{*N}}\left(e^{|\alpha|^2}e^{A(\alpha^*a+\alpha b)}\right)|
_{\alpha=\alpha^{*}=0},\nonumber\\
&=\frac{1}{N!}
\frac{\partial^{N}}{\partial\alpha^{*N}}[(\alpha^{*}+Ab)^{N}e^{Aa\alpha^{*}}]|_{\alpha^{*}=0},\nonumber\\
&=\frac{1}{N!}\sum_{m=0}^{N}\binom{N}{m}\left(\frac{\partial^{m}}{\partial\alpha^{*m}}(\alpha^{*}+Ab)^{N}\right) \nonumber \\
&\times \left(\frac{\partial^{N-m}}{\partial\alpha^{*N-m}} e^{Aa\alpha^{*}}\right)|_{\alpha^{*}=0},\nonumber\\
&=\frac{1}{N!}\sum_{m=0}^{N}\binom{N}{m}\frac{N!}{(N-m)!}A^{2(N-m)}(ab)^{N-m},\nonumber\\
&=\frac{1}{N!}\sum_{m=0}^{N}\binom{N}{m}\frac{N!}{(N-m)!}(-8i\beta\sin{\frac{\omega t}{2}})^{2(N-m)}.
\end{align}

\appendix{APPENDIX C: HAMILTONIAN CONTEXT}\label{a.H_Context}

For an explicit example of context for the generic Hamiltonian (\ref{e.H_o}) we provide a version of the original treatment by Jaynes \cite{Jaynes}. In this case we are describing a version of cavity or circuit QED. The electromagnetic field mode is interpreted as an oscillator, and the associated field Hamiltonian is written in terms of real dynamical variables, momentum $\hat{p}$ and co-ordinate $\hat{q}$ as:
\beq
\hat{H}=\frac{1}{2}\sum_{m}(\hat{p}_{m}^{2}+\omega_{m}^{2}\hat{q}_{m}^{2}).
\eeq
where $\hat{p}_{m}$ and $\hat{q}_{m}$ are the expansion coefficients of the vector potential and the transverse electric field:
\beqa
\hat{\vec{A}}(\vec{r},t)&=&-\sqrt{\epsilon_{0}}\sum_{m}\vec{u}_{m}(\vec{r}_{m})\hat{q}_{m}(t),\nonumber\\
\hat{\vec{E}}(\vec{r},t)&=&\sqrt{\epsilon_{0}}\sum_{m}\vec{u}_{m}(\vec{r}_{m})\hat{p}_{m}(t),
\eeqa
or their equivalents in a resonant circuit. In QED these are operators obeying canonical commutation relations. The function $\vec{u}_{m}(\vec{r}_{m})$ is the orthonormal vector function (i.e., incorporating polarization) that defines the $m^{th}$ mode of the field, which has frequency $\omega_{m}$.

The interaction Hamiltonian is derived in the familiar way from the dipole -field coupling: $\hat{H}_{int}=-\hat{\vec{d}}.\hat{\vec{E}}$. For a single excited mode labeled $a$, and a qubit with transition dipole matrix element $\vec{d}_{12}$, this takes the explicit form
\beqa
\hat{H}_{int}&=&-(\vec{d}_{12}.\vec{u}_{a}(0))\sqrt{\epsilon_{0}}\hat{p}_{a}(t)\hat{\sigma}_{x},\nonumber\\
&\equiv&\hbar\lambda(\hat{a}+\hat{a}^{\dagger})\hat{\sigma}_{x},
\eeqa
where we have introduced the coupling constant $\lambda$ used throughout, as well as the oscillator amplitudes via $\hat{p}_{a}=\sqrt{\hbar\omega/2}(\hat{a}+\hat{a}^{\dagger})$ and $\omega_{a}\hat{q}_{a}=i\sqrt{\hbar\omega_{a}/2}(\hat{a}-\hat{a}^{\dagger})$. As usual, the Pauli matrix expresses the operator character of the two-level dipole moment. In Sec \ref{s.comparison}, we removed the hats and regarded $\hat{p}_{a}$, $\hat{q}_{a}$, $\hat{a}$, and $\hat{a}^{\dagger}$ as classical numerical variables.

\end{document}